\documentclass[twocolumn]{revtex4}
\usepackage{xcolor}
\usepackage{array}

\usepackage{amsmath, amsfonts, amssymb, graphicx, color,xfrac}

\newcommand{\TT}{}
\graphicspath{{./figures/}}

\newcommand{\av}[1]{\langle{#1}\rangle{}}
\newcommand{\<}{\langle}

\renewcommand{\>}{\rangle}
\newcommand{\dotF}{\dot{\cal F}}

\newcommand{\revision}{}
\newcommand{\beq}{\begin{equation}}
\newcommand{\eeq}{\end{equation}}

\newcommand{\lastequal}{Corresponding authors. These authors contributed equally.}

\begin{document}

\newcommand{\deftitle}{{Fast decisions with biophysically constrained
gene promoter architectures}}

\title{\deftitle}

\author{Tarek Tohme}
\affiliation{Laboratoire de physique de l'\'Ecole normale sup\'erieure,
  CNRS, PSL University, Sorbonne Universit\'e, and Universit\'e de
  Paris, 75005 Paris, France}
  \author{Massimo Vergassola}
\affiliation{Laboratoire de physique de l'\'Ecole normale sup\'erieure,
  CNRS, PSL University, Sorbonne Universit\'e, and Universit\'e de
  Paris, 75005 Paris, France}
\author{Thierry Mora}
\thanks{\lastequal}
\affiliation{Laboratoire de physique de l'\'Ecole normale sup\'erieure,
  CNRS, PSL University, Sorbonne Universit\'e, and Universit\'e de
  Paris, 75005 Paris, France}
\author{Aleksandra M. Walczak}
\thanks{\lastequal}
\affiliation{Laboratoire de physique de l'\'Ecole normale sup\'erieure,
  CNRS, PSL University, Sorbonne Universit\'e, and Universit\'e de
  Paris, 75005 Paris, France}

\begin{abstract}
Cells integrate signals and make decisions about their future state in short amounts of time. A lot of theoretical effort has gone into asking how to best design gene regulatory circuits that fulfill a given function, yet much less is known about the constraints that  performing that function in a small amount of time imposes on circuit architectures.  Using an optimization framework, we explore the properties of a class of promoter architectures that distinguish small differences in transcription factor concentrations under time constraints. We show that the full temporal trajectory of gene activity allows for faster decisions than its integrated activity represented by the total number of transcribed mRNA. The topology of promoter architectures that allow for rapidly distinguishing low transcription factor concentrations results in a low, shallow, and non cooperative response, while at high concentrations, the response is high and cooperative. In the presence of non-cognate ligands, networks with fast and accurate decision times need not be optimally selective, especially if discrimination is difficult. While optimal networks are generically out of equilibrium, the energy associated with that irreversibility is only modest, and negligible at small concentrations. Instead, our results highlight the crucial role of rate-limiting steps imposed by biophysical constraints.

\end{abstract}

\maketitle

\section{Introduction}

Cellular decisions about fate and phenotypes are encoded in the states of regulatory biochemical networks. These networks sense the state of the environment, often encoded by concentrations of signaling molecules such as morphogen gradients in development~\cite{Driever1988,Mosby2024, Christian2012, Briscoe2015}, antigens in an immune response~\cite{Feinerman2008a, Altan-Bonnet2005, Francois2013, Achar2022, Altan-Bonnet2020} or sugars and aminoacids in chemotaxis~\cite{Berg1977, Macnab1972a, Sourjik2012, Janetopoulos2008, Tu2013}, and respond by expressing subsets of genes. The ability of cells to precisely and rapidly estimate and respond to these concentrations  is tied to the architecture of the readout circuits: promoters and signaling cascades. This readout occurs in an environment with high intrinsic noise~\cite{Elowitz2002, Tkacik2011}. A lot is known about the molecular details of specific promoter architectures and optimal circuits for precision, robustness, evolvability and discrimination potential have been explored to understand the constraints that these functions bring~\cite{Tostevin2007, Tkacik2008b,Lan2011, Tkacik2011a, Lan2012, Desponds2020, Lammers2023}. However, less work has been concerned with how the ability to reliably readout concentration can be constrained by limited time. 

Time matters for development, immune response and chemotaxis. Cells commit to different transcriptional programs through a set of decisions about which genes to up and down regulate. However, they do not have a lot of time to measure the environmental cues that provide them with information needed to make these decisions. The design of the readout algorithms is further constrained by the cellular hardware: transcription factors and gene readout complexes have to arrive at the activation sites and assemble -- transcription and translation take time. 

Exploring different functions has identified many molecular trade-offs. Inspired by early work of Berg and Purcell~\cite{Berg1977}, previous work has asked about the constraints that  precision of gene expression imposes on molecular designs~\cite{Berg1977, Bialek2005a, Gregor2007, Mora2019b, Carballo-Pacheco2019,Singh2020, Singh2017, Malaguti2019, Kaizu2014,Sartori2015, Sartori2011,Eldar2003, Endres2009b, Endres2009,  Mora2010, Hu2010a, Lammers2023,Tran2018, Estrada2016}. Others~\cite{Tostevin2007, Saunders2009} have considered the limits that biochemical noise imposes on the readout of positional gradients, leading to the existence of optimal gradient length-scales that maximize readout precision.  Bialek and co-workers~\cite{Tkacik2008b, Petkova2016} proposed the principle of information maximization between the input (the maternally laid Bcd gradient) and the output (the gap genes), in order to explain the precise, experimentally measured expression profile of Hunchback~\cite{Gregor2007} and then other gap genes~\cite{Petkova2016}. Others~\cite{Okabe-Oho2009, Erdmann2009} further explored the idea suggested by Gregor et al.~\cite{Gregor2007} of spatial averaging of the protein concentrations as a mechanism of noise reduction. As an alternative to fixed-time decision making, Wald's Sequential Probability Ratio Test (SPRT)~\cite{Wald1945} applied to biochemical networks was used to show that receptors~\cite{Siggia2013} and gene regulatory networks~\cite{Desponds2020} can reach large precision in a shorter time than allowed by Berg-Purcell~\cite{Berg1977, Gregor2007, Kaizu2014} or maximum likelihood estimators~\cite{Endres2009b, Endres2009,  Mora2010}.

These examples show us that we can imagine different functional pressures acting on promoters. Different encoding of the signal can lead to different optimal designs, even if we impose the same objective: a fast and precise readout. Here we explore the question whether different encoding of the signal results in different optimal promoter architectures, {\revision assuming the signal is decoded in an optimal way}. What constraints does limited readout time impose on molecular implementations of transcription? Specifically, focusing on the simplest gene regulatory networks that can function out of equilibrium, we ask {how much faster decisions can be made if the whole series of binding and unbinding events is exploited rather than the accumulated product of the transcription process, and how that distinction as well as physical constraints affect the architecture of the optimal network.}

{\revision In the same spirit as the Berg and Purcell scheme~\cite{Berg1977}, and the line of work descending from it~\cite{Bialek2005a, Gregor2007, Mora2019b, Carballo-Pacheco2019,Singh2020, Singh2017, Malaguti2019, Kaizu2014,Sartori2015, Sartori2011,Eldar2003, Endres2009b, Endres2009,  Mora2010, Hu2010a, Lammers2023,Tran2018, Estrada2016}, we adopt the point of view which separates finding the physical limits under which biology operates from identifying the mechanisms it may use to reach these limits. Practically, to obtain performance bounds that no physical scheme can surpass, we study an idealized observer that has access to the full history transcription, with no assumption about whether such an observer is actually realized biologically.
Whether regulatory networks decode using the full activity, or an incomplete, distorted, cumulative, or otherwise processed signal, remains an important question. If the process is encoded transcriptionally, the first stage of this information processing pipeline is the encoding of concentrations in mRNA transcription activity, which is the object of our study.}

{\revision We start by unifying previous works that interpret the readout of a biochemical network as an on-the-fly decision \cite{Desponds2020, Lammers2023, Siggia2013}, and those that view it as a fixed-time estimate of concentration \cite{Berg1977, Gregor2007, Kaizu2014, Endres2009b, Endres2009,  Mora2010} under the common framework of Fisher information (first two sections of the results). We further highlight the differences in decoding with accumulated and full-trace evidence. We then study optimal architecture for a single ligand and a single binding site, and generalize to multiple binding sites and to the additional presence of spurious ligands. We end with a discussion of the energetic costs associated with optimal networks.}

\section{Results}

\subsection{Decision time in a gene regulation model}~\label{SPRT}

\begin{figure*}
  \centering
		\includegraphics[width=\textwidth]{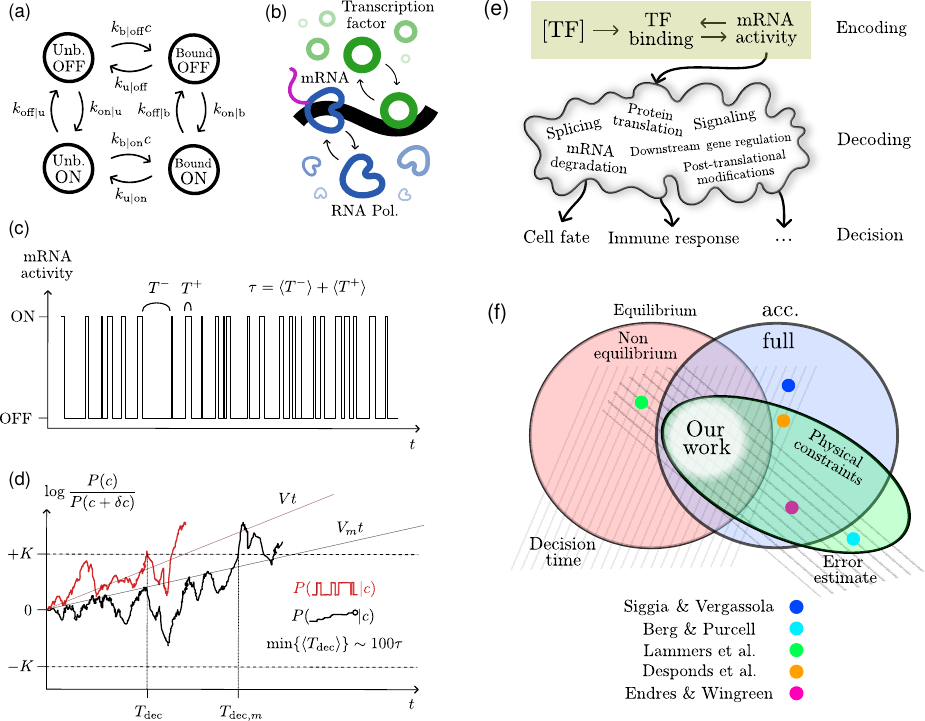} 
		{\caption{
		(a) A model of a receptor to which a ligand can bind and activate a biochemical response (ON/OFF). Binding activity is coupled to the response through 8 transition rates. (b) A promoter with one binding site regulating a gene, an example of a system described by (a). (c) Trace of transcription activity over time. (d) Log likelihood ratio for a given activity trace for two readout mechanisms: full activity (red trace) and cumulated activity decoding (black trace), the stochastic SPRT decision times $T$ and $T_m$, and the SPRT drift terms $Vt$ and $V_mt$ associated with each scheme. Minimum decision times for the circuit in (a) are on the order of 100 activity cycles. \textcolor{blue}{(e) A schematic view of the information flow from a transcription factor concentration all the way to a biochemical decision. Our work's scope is at the level of mRNA transcription activity, focusing on the best possible decoding from that stage. (f) A visual literature review highlighting key connections to previous work.}
		\label{4circuit}}}
\end{figure*}

We consider a simple non-equilibrium model of gene regulation, where the promoter may be active (on), meaning that it gets bound by RNA-polymerase to make transcripts, or not active (off). It may also be bound or not by a transcription factor (Fig.~\ref{4circuit}b). This results in a four-state system summarized in Fig.~\ref{4circuit}a. The binding rates $k_{\rm b|off}c$, $k_{\rm b|on}c$ of the transcription factor (TF) onto the DNA scale linearly with the TF concentration $c$. They also depend on the activity state on/off, as do the unbinding rates $k_{\rm u|off}$, $k_{\rm u|on}$. Likewise, the activation $k_{\rm on|b},k_{\rm on|u}$ and deactivation $k_{\rm off|b},k_{\rm off|u}$ rates depend on the binding state. Later we shall consider generalizations of this model to multiple binding sites for the transcription factor, as well as the possibility of non-specific binding.
Unless otherwise specified, we assume that the system has reached a steady state.

We ask how long an ideal observer of the sequence of activity would need to distinguish between two similar concentrations $c$ and $c+\delta c$ of the transcription factor regulating the gene \cite{Siggia2013,Desponds2020}. While there is no guarantee that cells can implement decisions as well as an ideal observer, this puts a hard physical bound on the precision of any process downstream of gene expression (Fig.~\ref{4circuit}e). Formally, the ideal observer has access to a temporal trace of transcriptional activity  (gray lines in Fig.~\ref{4circuit}c), which can be encoded by a sequence of times spent in the active and inactive states up to time $t$, $O(t)=(T^-_1,T^+_1,T^-_2,\ldots)$. For each concentration $c$, one can define the log-likelihood of these observations given $c$:
\beq
\mathcal{L}(c;t)=\ln P(O(t)|c).
\eeq
The Sequential Probability Ratio Test (SPRT), originally proposed by Wald~\cite{Wald1945}, consists of tracking the log-likelihood ratio between the two competing hypotheses ($c$ and $c+\delta c$):
\beq\label{langevin}
\mathcal{R}(t)=\log\frac{P(O(t)|c)}{P(O(t)|c+\delta c)}=\mathcal{L}(c;t)-\mathcal{L}(c+\delta c;t).
\eeq
When the true concentration is $c$, $R(t)$ follows a stochastic process that is biased towards positive values, reflecting the accumulation of evidence from the observations (Fig.~\ref{4circuit}d). A decision is made when evidence has reached some threshold of certainty, $|{\cal R}(t)|>K\equiv \log ((1-\varepsilon)/\varepsilon)$, where $\varepsilon$ is the error level, {\revision which we fix in all numerical experiments at $\int_{|x|>1}dx/\sqrt{2\pi}e^{-x^2/2}=0.32$, the probability of making an estimate with an error greater than one standard deviation, which is well approximated by a Gaussian distribution in the hard decision regime \cite{Lammers2023, Desponds2020}}. 
The observer determines that the concentration is $c$ if ${\cal R}(t)>K$ and $c+\delta c$ if ${\cal R}(t)<-K$. The time of this decision $T_{\rm dec}$, is itself a random variable that depends on the stochastic trajectory of ${\cal R}(t)$.

The smaller $\delta c$, the harder the two hypotheses are to distinguish, and the longer it takes for that evidence to accumulate. In that limit, and under reasonable assumptions that the active and inactive interval durations are not correlated over long times, the law of large numbers allows us to approximate the evolution of the log-likelihood ratio as a random walk \cite{Siggia2013}:
\beq
\frac{d \mathcal{R}}{d t} = V +\sqrt{2 V} \eta, \label{V_def}
\eeq
where $\av{\eta(t) \eta(t')}\sim \delta (t-t')$. This equation may be integrated explicity, so that $R(t)$ is normally distributed with mean $Vt$ and variance $2Vt$. Note that the diffusion term is equal to the drift in \eqref{V_def}, which is made necessary by the condition $\<e^{-\mathcal{R}}\>=\sum_O P(O|c)P(O|c+\delta c)/P(O|c)=1$ \cite{Desponds2020}.

Assuming that $c$ is the true concentration, $V$ is:
\beq
V=\lim\limits_{t \rightarrow \infty} \frac{\left \langle \mathcal{R}(t) \right \rangle_c}{t}=\lim\limits_{t \rightarrow \infty} \frac{D_{\rm KL}[P(O(t)|c) ||P(O(t)|c + \delta c)]}{t} , \label{V_lim}
\eeq
where $\<\cdot\>_c$ denotes an average over $P(O|c)$, and $D_{\rm KL}$ is the Kullback-Leibler divergence. 
In the limit where decisions are hard, $\delta c\ll c$, we may expand this expression at small $\delta c$:
\beq
V\approx \frac{1}{2}{\left(\frac{\delta c}{c}\right)}^2 \dot{\cal F}(c),
\eeq
where
\beq\label{dotF}
\dot{\cal F}(c)=- \lim\limits_{t \rightarrow \infty} \frac{1}{t} \left\<\frac{\partial^2 {\cal L}(c)}{(\partial \ln c)^2}\right\>_c,
\eeq
is the Fisher information rate for the log-concentration. 
Even for a simple model such as in Fig.~\ref{4circuit}, that rate does not have a general analytical expression. We developed a numerical procedure based on a combination of Monte-Carlo sampling and dynamic programming to compute $\dotF(c)$ efficiently for any Markov system with active states (see Section~\ref{sec:likelihood}).

The decision time is given by the mean first-passage time for this biased random walk in the log-likelihood space with decision boundaries set by $K$. For $\delta c\ll c$, it is~\cite{Siggia2013, Desponds2020}: 
\beq
\langle T_{\rm dec}  \rangle= {\frac{K}{V} \tanh \left(\frac{K}{2} \right) }.
\label{MFPT}
\eeq
Large deterministic bias $V$ reduces the decision time (Fig.~\ref{4circuit}d). This result is general and does not depend on the studied problem. In the limit of low errors $\varepsilon \to 0$, $K\rightarrow \infty$, and $\langle T \rangle_{\rm dec} \sim K/V$.

Estimates of the decision time \eqref{MFPT} assume full knowledge of the activity trace. Alternatively, one may assume that observations are incomplete or reduced. For instance, Lammers et al~\cite{Lammers2023} studied SPRT where observations were reduced to the cumulated time of active transcription, $T_{\rm on}(t)=\sum_{i}T_i^+$. Doing so leads to an expression for the reduced Fisher information rate $\dot{\cal F}_m(c)$, obtained using the same Eq.~\ref{dotF}, but using $\mathcal{L}_m(c;t)=\ln P(T_{\rm on}(t)|c)$ instead of ${\cal L}=\ln P(O(t)|c)$. While it only provides a lower bound to the full information rate, Lammers et al.~\cite{Lammers2023} showed that it can be computed analytically as:
\beq\label{Fm}
\dot{\cal F}_m(c)=\frac{s^2}{\dot v}\leq \dot{\cal F}(c),
\eeq
where
\beq
s=c\frac{\partial p}{\partial c}
\eeq
is the logarithmic sensitivity of the activation curve $p(c)\equiv  P({\rm on}|c)=\lim_{t\to\infty} T_{\rm on}(t)/t$, and
\beq\label{vdot}
\dot v=\lim\limits_{t\to\infty}\frac{1}{t}\mathrm{Var}(T_{\rm on}(t))
\eeq
is the rate of increase of the variance of the cumulated activity time. Both $s$ and $\dot v$ can be computed from the transition rates of the model, using asymptotic formulas for Markov processes \cite{Whitt1992}.

We will compare this partial readout mechanism with the one based on the full likelihood, $P(O|c)$. We will show not only that the latter extracts more information about concentration per unit time, leading to faster decisions under SPRT, but also that minimizing decision time under one decoding scheme or the other yields different optimal gene activation circuits. This highlights how conclusions drawn from information-optimization arguments may depend intricately on the nature of the available information, the details of the functions being optimized, and under which constraints. We discuss these aspects in the next sections.

\subsection{Link to the accuracy of concentration sensing}
In their pioneering paper, Berg and Purcell calculated the accuracy with which a receptor can sense the concentration $c$ of a freely diffusing ligand within a fixed time $T$~\cite{Berg1977}. Here we show how to link these accuracy bounds to the SPRT framework for decision times.

Within our framework, the question of accuracy may be addressed by computing the error made when estimating $c$ from the on/off observations $O(t)$ using maximum likelihood, $\hat c(t)=\mathrm{arg}\max_c P(O(t)|c)$. At long times, the scaling of the error $\delta \hat c=\hat c(t)-c$ is given by the Cram\'er-Rao bound, which states that it is governed by the Fisher information rate:
\beq\label{accuracy}
\frac{\<\delta \hat c^2\>}{c^2}\approx \frac{1}{{\revision T} \dot{\cal F}(c)}.
\eeq
As a result, both the SPRT decision time, and the accuracy of concentration estimate at fixed time, are governed, in the limit of hard decision and long times, by the same quantity: the Fisher information rate $\dot{\cal F}(c)$. {\revision In other words, making an accurate estimate of concentration on the one hand, and making a fast binary decision between two competing hypotheses on concentration on the other, while they are different tasks (making an estimate at fixed time versus making a decision on the fly at variable time) best carried out by different decoding schemes, correspond to the same encoding objective. Having made this link, the rest of the paper adopts the Fisher rate as the quantity of interest, regardless of whether the readout is used to make a decision between two concentrations or to sense a concentration with high precision.} In the following sections we shall focus on optimizing that quantity for various choices and constraints on the gene regulation model.

Classical results about readout accuracy \cite{Berg1977,Endres2009} may be recovered using \eqref{accuracy}. 
We can solve analytically a particular case of our simple gene regulatory model, in which the gene is active only when bound, and inactive when unbound. That case is achieved by considering the limit $k_{\rm on |b}, k_{\rm off|u}\to\infty$, and $k_{\rm u|off}=0$, $k_{\rm b|on}=0$. The system then makes transitions from an inactive, unbound to an active, bound state, and back, with effective rates $k_{\rm b}\equiv k_{\rm b|off}$ and $k_{\rm u}\equiv k_{\rm u|on}$. This model of a 2-state system has been widely used, beyond the context of gene regulation, to model the binding of ligands by receptors.
The Fisher information rate may be computed analytically \cite{Endres2009}:
\beq\label{2state}
\dot{\cal F}(c)=(1-p(c))k_{\rm b}c, 
\eeq
where $p(c)=(1+k_{\rm u} /k_{\rm b}c)^{-1}$ is the probability of being active. Note that this rate is exactly the average number of binding events per unit time, with the $1-p$ term ensuring that the binding site must be unoccupied for a binding event to occur. Further assuming that the binding rate is diffusion limited, with the binding site modeled as a circle of radius $a$, and the transcription factor having diffusivity $D$, we have $k_{\rm b}=4Da$ and:
\beq\label{eq:endres}
\frac{\<\delta \hat c^2\>}{c^2}\approx \frac{1}{4Dac(1-p)t},
\eeq
which is the result of \cite{Endres2009}.

The reduced information rate, where only the total bound time is used, may also be computed exactly using \eqref{Fm}, yielding
$\dot{\cal F}_m(c)=(1-p)k_{\rm b}c/2$ \cite{Lammers2023}, which is exactly half of the full rate. This leads to an increased error in the estimate of the concentration at fixed time, $\hat c_m(t)=\mathrm{arg}\max_c \mathcal{L}_m(c,t)$:
\beq\label{BP}
\frac{\<\delta \hat c_m^2\>}{c^2}\approx \frac{1}{2Dac(1-p)t},
\eeq
which is exactly the result obtained by Berg and Purcell for a single receptor \cite{Berg1977}.

In summary, $\mathcal{F}$ and $\mathcal{F}_m$ correspond to two estimators previously discussed in the literature: maximum likelihood and mean occupancy.
The estimator $\hat c_m(t)$ is in fact obtained by inverting the relation $T_{\rm on}(t)/t=p(c)$, which is the decoding scheme originally studied by Berg and Purcell. The twofold difference in accuracy between using $\hat c$ versus $\hat c_m$, which was first noted in \cite{Endres2009}, is here reframed in terms of using the likelihood of the full activity trace, versus the average activity. Physically, the difference stems from the fact that the information rate $\dotF$ only relies on unbinding times to discern concentrations, while the reduced information rate $\dotF_m$ is subject to noise in both the bound and unbound time periods, the former containing no information about concentration in the 2-state system. This doubles the sources of uncertainty, as discussed in detail in Refs~\cite{Endres2009,Carballo-Pacheco2019}. {\revision The relation of our results to previous literature is summarized in Fig.~\ref{4circuit}f.}

{\revision Since in this section we have shown that the following definitions are equivalent,  in the rest of the paper we will use the terms `optimal' or `highest precision' to mean: i) the minimal error at a given mean decision time; ii) the shortest mean decision time for given error level; iii) the maximum Fisher information rate.}

\subsection{Different optimal encodings at high and low concentration}
\label{results_rate_constraints}

\begin{figure*}
  \centering
	\includegraphics[width=0.99\linewidth]{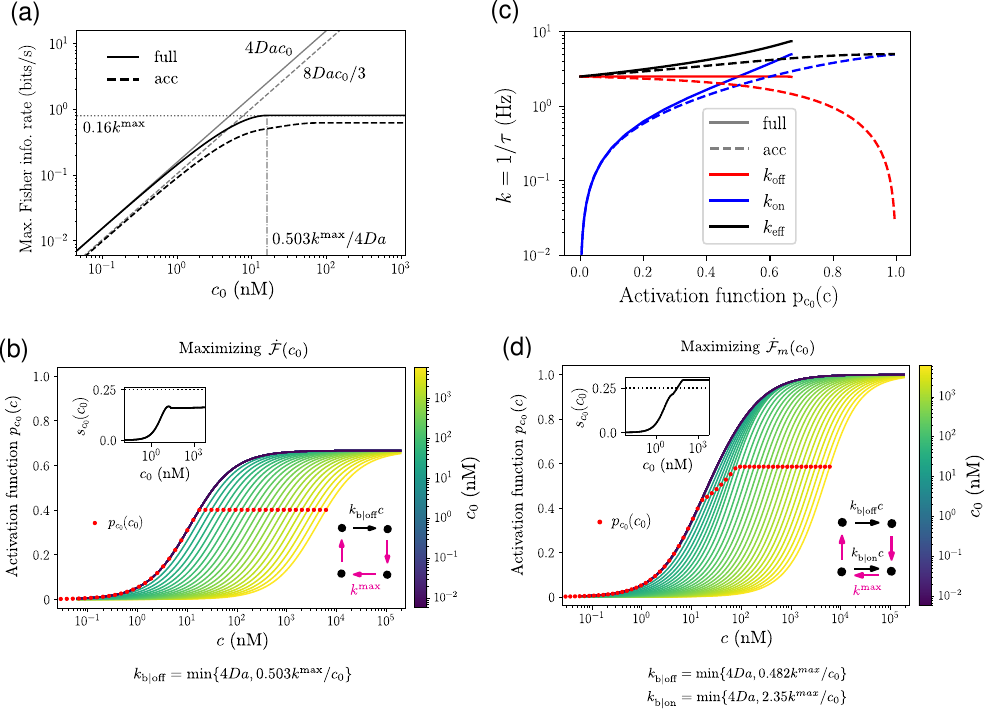}
	{\caption{
            Minimizing decision time under diffusion, activation and unbinding time limits. (a) Optimal Fisher information rates at different concentrations. $D$ is the diffusion constant of the ligand in $\mu$m$^2$s$^{-1}$, $a$ is the length of the receptor in $\mu$m, giving $4Da$ units of ${\rm M}^{-1}{\rm s}^{-1}$. $k^{\rm max}$ is the maximum activation rate in ${\rm s}^{-1}$. {\revision Throughout the text, labels `full' and 'acc' refer respectively to the full trace and accumulated mRNA decoding scheme.} (b) Activation functions $p_{c_0}(c)$ of a set of circuits maximizing $\dot{{\cal F}}(c_0)$ at different concentrations $c_0$. Inset: sharpness of the activation function $s_{c_0}(c_0) = c_0\partial p_{c_0}(c)/\partial c|_{c_0}$ at the true concentration. Dotted line at 0.25 indicates a Hill coefficient of 1. The decrease compared to the inset of (d) shows that $\dot{{\cal F}}$-optimal circuits do not optimize for sharpness at high concentration. {\revision The architecture of optimal circuits is shown on the lower-right corner in (b) and (d). States and arrows follow the layout of Fig. \ref{4circuit}a. Arrows are colored black if their rate equals $k_{\rm b | off}$ and purple if it equals $k^{\rm max}$. Arrows not shown saturate $k_{\rm min}$}. (c) Effective rates of {\revision on ($k_{\rm on}=1/\tau_{\rm on}$, blue) and off ($k_{\rm off}=1/\tau_{\rm off}$, red)} activity cycles as a function of mean activity $p_{c_0}(c)$ of a set of circuits optimized at $c_0 = 3.36\;\rm nM$, evaluated at different values of $c$. Black lines indicate the correlation time between bursts, $k_{\rm eff} = 1/\tau_{\rm on} + 1/\tau_{\rm off}$. (d) Equivalent figure for $\dot{{\cal F}}_m$-maximizing circuits.
			\label{difflimitted_simple}}}
\end{figure*}

We now set out to find the circuits that best discriminate between transcription factor concentrations around $c=c_0$. This translates into finding the set of transition rates that maximize $\dot{{\cal F}}(c_0)$, or alternatively $\dotF_m(c_0)$ if one assumes the decoder can only use the cumulated activity.

Decisions can be sped up by increasing indefinitely all the rates to very high values. However, physics imposes constraints on the rates. Binding events are often limited by the diffusion time of the ligand. Diffusion rates in turn are limited by the cellular environment and have accurately measured bounds, e.g., in the context of early fly development~\cite{Gregor2007a, Abu-Arish2010, Porcher2010, Tran2018}. We estimate upper bounds for unbinding rates from the minimum bound time required to trigger a downstream mechanism, such as  activating a gene (which is an assumption made in Ref.~\cite{Desponds2020}). As a result, in a realistic system, all four states have non-zero occupancy probability. In addition to making the models more physically realistic, these constraints play a central role in determining the optimal encoding resulting from each of the two readout mechanisms we consider.

Throughout the text, unless otherwise noted, optimizations are done under the constraints imposed by ligand diffusion and the minimum time for activation and unbinding involved in transcription. Diffusion sets the upper bound on binding factors: $k_{\rm b|on,off} \leq 4Da$. All other concentration-independent rates are upper-bounded by $k_{\rm max}$. We also impose a general lower limit $k_{\rm min}$ on all rates for numerical stability.

The maximization of the information rate under these constraints and at a given $c_0$
reveals two distinct behaviours in the low and high concentration regimes, as illustrated in Fig.~\ref{difflimitted_simple}a for both $\dotF$ and $\dotF_m$: first a linear increase of information with $c_0$, then a plateau at high $c_0$.

The circuits maximizing $\dotF(c_0)$ all have the same cyclic structure (Fig.~\ref{difflimitted_simple}b, lower inset). {\revision These structures are strongly out of equilibrium, as we will discuss further in Section \ref{results_noneq}}.
Once bound, the system traverses the (bound, on) state and (unbound, on) states to go back to the starting point (unbound, off) as quickly as possible, maximizing all the corresponding rates at their upper bound, while all reverse rates saturate the lower bound. {\revision Within the constraints on rates, this circuit approaches the most closely the 2-state system achieving the bound of Eq.~\ref{eq:endres}}.
The sharp transition observed in Fig.~\ref{difflimitted_simple}a is driven by a sharp transition in the 
optimal binding rate $k_{\rm b|off}$ as a function  of $c_0$. When ligands are relatively scarce (small $c_0$), the waiting time for binding an inactive gene is the rate limiting step, and the binding rate saturates the diffusion-limited bound, $k_{\rm b|off}=4Da$.
When ligands are abundant (large $c_0$), TF are no longer constrained by the diffusion limit. However, the optimal binding rate $k_{\rm b|off}c_0$ saturates to about half of the maximum value of the non-binding rates, $k_{\rm b|off}c_0\approx 0.503 k_{\rm max}$ (see Appendix~\ref{App:V_limits_4s} for a derivation of the prefactor, which is the solution of a transcendental equation). The Fisher information rate is then $\dot{{\cal F}} \approx 0.16 k_{\rm max}$; that limit is plotted as a horizontal line in Fig.~\ref{difflimitted_simple}a. The reason for this saturation is that, while the optimal circuit seeks to complete binding and activation cycles as quickly as possible, it also tries to remain as sensitive as possible to small changes in concentration, and limits its binding rate to avoid being always active.
In sum, we have $k_{{\rm b|off}}(c_0)=\min(4Da, 0.503 k_{\rm max}/c_0)$, with the transition between the two regimes happening when $c_0\approx 0.503 k_{\rm max}/4Da$. {\revision To summarize, the optimal binding rate is strictly determined by the physical constraints on diffusion and activation time. The latter dominates at high concentration, while the binding rate saturates the former at low concentration.}

Across both regimes, the gene activation curve of the network optimized for discriminability around $c_0$ (Fig.~\ref{difflimitted_simple}b) is:
\beq
p_{c_0}(c)=\frac{2}{3+k_{\rm max}/(k_{{\rm b|off}}(c_0)c)}.
\eeq
The curves all saturate at $\sim 2/3$, as the network spends a third of its time in the inactive, bound state as $c\to\infty$. Since binding only occurs in the inactive state, only the mean inactive time, $\tau_{\rm off}=1/4Dac+1/k_{\rm max}$, depends on concentration, while the mean active time, $\tau_{\rm on} =2/k_{\rm max}$, is a constant (Fig.~\ref{difflimitted_simple}c).
In the low concentration limit ($4Dac_0\ll k_{\rm max}$), the optimal circuit, operating at the concentration $c_0$ for which it was optimized, spends most of its time in the unbound, off state, $p_{c_0}(c_0)\approx  8Dac_0/k_{\rm max}$.
Since ligands are limited, the system tries to process each binding event as quickly as possible, to make itself available to the next one.
In effect, this strategy emulates the 2-state system considered previously (Eq.~\ref{2state}), with effective off rate $k_{\rm max}/2$. The optimized information rate yields the same value as in that system, $\dotF(c_0)\approx 4aDc_0$, plotted as a gray solid line in Fig.~\ref{difflimitted_simple}a. 
When $Dac_0>0.503 k_{\rm max}$, the optimal expression sets on $p_{c_0}(c_0)\approx .401$ (red dots in Fig.~\ref{difflimitted_simple}b).  Note that optimal networks in general do not try to maximize the sensitivity $s=dp_{c_0}/dc(c_0)$, which remains small even at large $c_0$ (inset of Fig.~\ref{difflimitted_simple}b).

The optimal circuits that maximize the reduced Fisher information rate $\dotF_m(c_0)$ are markedly different from those that maximize the full rate $\dotF(c_0)$ (Fig.\ref{difflimitted_simple}d, lower inset). The key difference is that, over a wide range of concentrations, the binding rate while active maximizes the diffusion bound, $k_{\rm b|on}=4Da$. Intuitively, since the reduced observable $T_{\rm on}(t)$ is ``forced'' to read off the duration of active periods whether they are informative about $c$ or not, and is thus subject to their stochasticity, optimal circuits seek to make the best of it by packing information into those periods. Having a non-vanishing back rate $k_{\rm b|on}$ introduces a dependence on concentration of the duration of active periods, which circuits can exploit. As a result, the durations of active and inactive periods both depend on $c$ (Fig.~\ref{difflimitted_simple}c, blue and red dashed lines), but with opposite trends, so that the effective autocorrelation inverse time, $k_{\rm eff}=1/\tau_{\rm on}+1/\tau_{\rm off}$,  {\revision depends much more weakly on activation} (Fig.~\ref{difflimitted_simple}c, dashed black line), in contrast with $\dotF$-optimal networks (solid black line).

In the low concentration limit however, $4Dac_0\ll k_{\rm max}$, that back rate is negligible and the circuit behaves effectively like an irreversible cycle, just like the optimal circuit for $\dotF(c_0)$.
However, its reduced information rate is limited by the stochasticity of active periods, in a similar way to the 2-state system discussed earlier (Eq.~\ref{BP}). As a result, $\dotF_m(c_0)\approx (2/3)\dotF(c_0)$. Recall that, in the 2-state system, the loss factor relative to the full trace $\dotF$ was $1/2$. Here, it is only $2/3$ because the active periods are less stochastic than in the 2-state system, as they now involve two steps to deactivation: first unbinding, and then deactivation itself, each with rate $k^{\rm max}$. This implies $\<T^{+}\>=2/k^{\rm max}$ and $\<(\delta T^+)^2\>=2/(k^{\rm max})^2$, i.e. coefficient of variation $\<(\delta T^+)^2\>/\<T^+\>^2=1/2$, instead of 1 in the 2-state system. As a result, an observer of the cumulated activity is subject to the sum of fluctuations of the binding time, with coefficient of variation of 1, which is also shared by the observer of the full trace, and of the active time, with coefficient of variation of 1/2, which the observer of the full trace ignores since it contains no information on $c$. As a result, the total stochasticity in $\dot v$, defined by Eq. \eqref{vdot}, is $50\%$ larger than for an observer of the full activity trace, hence the $2/3$ factor in $\dotF$.
A more formal derivation is provided in Appendix~\ref{App:V_limits_4s}.

For higher concentrations, the system goes through two transitions as a function of $c_0$. First, for similar reasons as for $\dotF$, the binding rate while inactive $k_{\rm b|off}$ saturates a bound of $0.482k_{\rm max}/c_0$. Second, $k_{\rm b|on}$ knows the same fate but at a higher value, $2.35 k_{\rm max}/c_0$. Those saturations are necessary to keep the activity at $c_0$ in the range of reasonable sensitivity.

In contrast to the optimal circuits for $\dotF$, the response curves of the $\dotF_m$-optimal circuits (Fig.~\ref{difflimitted_simple}d) have the property $p_{c_0}(c)\to 1$ as $c\to\infty$: since both binding rates are non-vanishing, the system spends almost all its time in the (bound,on) state as $c\to\infty$. Mean gene activity $p_{c_0}(c_0)$ (red dots) and sensitivity $s(c_0)$ (inset) are both small at low concentrations $c_0$, and increase for larger ones.
In the large concentration regime, gene activity reaches a maximum $p_{c_0}(c_0)\approx 0.586$, and so does the sensitivity $s=0.295$. That sensitivity is larger than for the $\dotF$-optimal circuits, and also than the 1/4 of a classic activation curve obtained in a 2-state system (Hill with cooperativity of 1), meaning that these circuits exploit the 4-state architecture to implement an effective cooperativity. However, it is still smaller than the maximum sensitivity of 1/2 (Hill with cooperativity 2) reported in \cite{Lammers2023}.

To summarize, when ligands are limited, optimal circuits want to stay as inactive as possible to capture the maximum number of binding events. That result holds for both the full observations ($\dotF$) or for the cumulated activity ($\dotF_m$). {\revision In that regime, the only constraint driving the behaviour is the diffusion limit, all other constraints being essentially non-limiting.} When ligands are no longer limited, physical constraints on the rates of gene operation take over. The strategy for the full observation is markedly different than for the cumulated activity, favoring less active circuits with a more shallow response curve, and in which the duration of activity periods does not depend on concentration. By contrast, circuits optimized for cumulated activity exploit the duration of active periods to enhance sharpness and utilize the full range of possible activity levels.

\begin{figure*}
	\centering
	\includegraphics[width=0.9\linewidth]{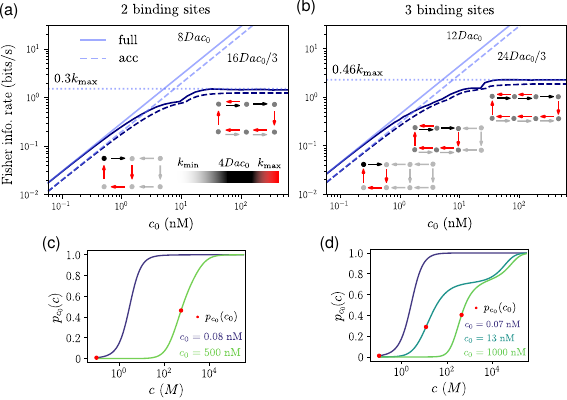}
	
	{\caption{
	Onset of cooperativity for optimal receptors with multiple binding sites. (a) Optimal Fisher information rates at concentrations ranging from $0.06$ nM to $600$ nM. Full trace decoding indicated by solid lines, and accumulated mRNA decoding by dashed lines. $D$ is the diffusion constant of the ligand in $\mu$m$^2$s$^{-1}$, $a$ is the length of the receptor in $\mu$m, giving $4Da$ units of ${\rm M}^{-1}{\rm s}^{-1}$. $k^{\rm max}$ is the maximum activation rate in ${\rm s}^{-1}$. Optimal promoter architectures for the full trace decoding scheme are shown as insets. {\revision Darker shading of states indicates higher occupation probability}. On the left hand side, optimal circuits take on a diffusion-limited, non-cooperative configuration where the two sites bind ligands independently, and receptors remain mostly unoccupied. On the right, the Fisher information rate is limited by activation time $1/k_{\max}$ and the two sites bind cooperatively. (b) Corresponding plot for three binding sites, showing two transitions to cooperativity. (c,d) {\revision Activation functions of optimal circuits for two and three binding site receptors, respectively, for circuits optimized at 2 concentrations $c_0$ corresponding to the regimes of cooperativity outlined in (a) and (b)}. \label{difflimitted_2bs}}}
\end{figure*}

\subsection{Transition to cooperativity with multiple binding sites}

The regulatory architecture coupling TF binding and gene activity can be generalized to multiple identical binding sites, by defining binding states corresponding to the number of bound TF, and corresponding binding rates: $k_{\rm 1b|off}$ for the first binding event, $k_{\rm 2b|off}$ for the second, etc., and likewise for the binding rates while active, as well as unbinding rates (Fig.~\ref{difflimitted_2bs}a). Binding rates are now limited by $4Da$ times the number of free binding sites, and similarly unbinding rates by $k_{\rm max}$ times the number of occupied binding sites.

We optimized $\dotF$ and $\dotF_m$ numerically with 2 binding sites, varying the reference concentration $c_0$ (Fig.~\ref{difflimitted_2bs}a). The optimization revealed 3 regimes, corresponding to effectively two simple network architectures  (insets).

For $c_0<c_0^*\approx 0.5 k_{\rm max}/8Da$, the optimal network allows for only one binding site to be occupied. Binding of either binding site happens with diffusion limited rate $2\times 4Da$, at which point the gene activated  is with the maximal rate $k_{\rm max}$, and then unbound and deactivated as fast as possible. In effect, this solution is equivalent to the single binding site solution in the diffusion limited regime, but with a doubled effective binding rate $8Da$ thanks to the existence of 2 binding sites. The optimal network for cumulated activity time behaves similarly in that limit, but with the same $2/3$ factor relative to the full trace, due to the added uncertainty of the duration of the active period.

For $c>c_0^*$, the optimal solution changes drastically to a network in which activity is triggered only once both binding sites have been bound. This design, in which both binding rates while inactive saturate their diffusion-limited bound, corresponds to a maximally cooperative response. A second transition occurs as $c_0$ is increased and the optimal binding rates cease to be diffusion limited, but instead scale with $k_{\rm max}$. The maximum Fisher information scales with $k_{\rm max}$.

The result generalizes beyond 2 binding sites. With 3 binding sites, the system undergoes 2 transitions towards cooperativity, from a just one binding site being sufficient to activate transcription, to 2 and then 3, as the reference concentration $c_0$ is increased (Fig.~\ref{difflimitted_2bs}b). We expect similar behaviours for any $k\geq 2$ binding sites: when diffusion limited, the optimal network does not implement cooperativity but rather activates whenever any of the binding site is occupied, with rate $k\times 4Da$. When the binding rate becomes comparable to the other rates, the system requires that several binding sites be occupied in order to start transcription, thereby increasing its cooperativity (Fig.~\ref{difflimitted_2bs} c and d).

\subsection{Circuits with fastest decision time do not maximize selectivity against decoys}
\label{results_spec}

\begin{figure*}
	\begin{center}
		\includegraphics[width=\textwidth]{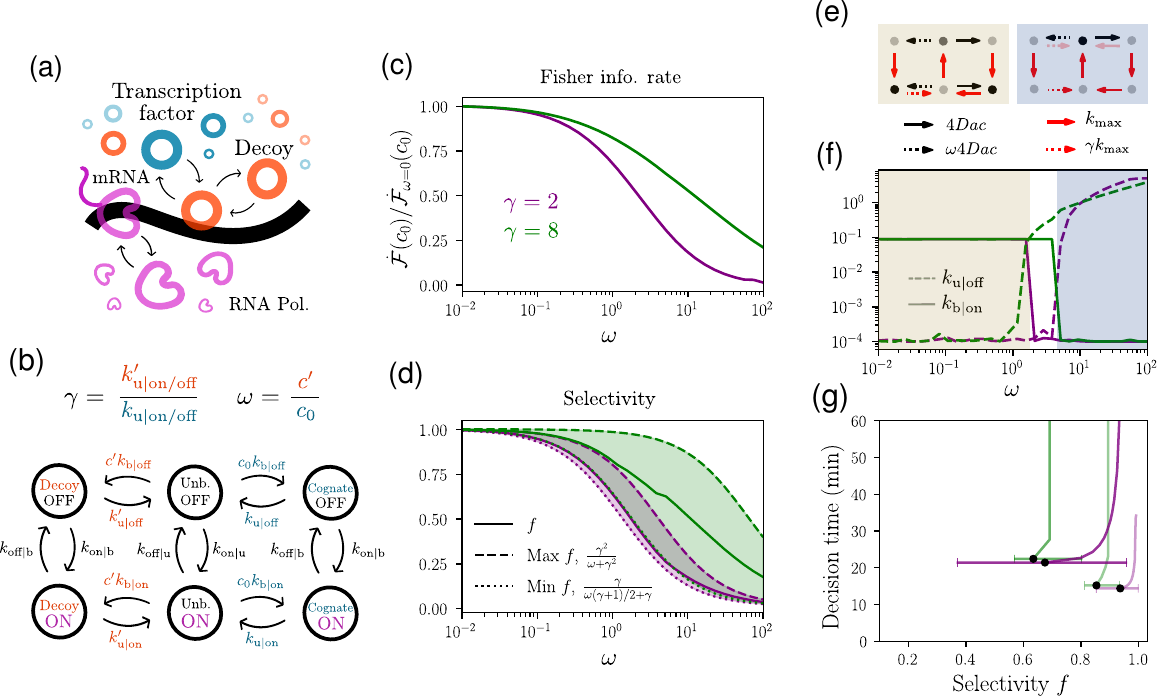}
		{\caption{ 
                Maximizing the Fisher information rate in the presence of decoy ligands. 
				(a) A promoter with a single binding site. (b) A general model for a receptor to which both cognate and decoy ligands can bind and activate a downstream response (ON/OFF). The factor $\gamma$ quantifies how much faster decoys unbind relative to cognate ligands, and $\omega$ is the ratio of their concentrations, respectively $c'$ and $c_0$. (c) Fisher information rate normalized by its value in a solution with no decoys, i.e. $\omega=0$, for specificities $\gamma\in\{2,8\}$ (d) Selectivity of the circuit with increasing $\omega$. Circuits  that make the fastest decisions in the presence of spurious ligands are neither minimally selective nor maximally selective against the decoys. (e) The two different architectures of optimal circuits in the presence of decoys. {\revision Arrows under each circuit indicate transition rates: diffusion-limited in black and limited by activation/deactivation speed in red. Light red indicates that rates are near the limit. For states, darker shading indicates higher occupation probability}. 
				(f) Two key transition rates $k_{\rm u|off}$ and $k_{\rm b|on}$ characterize the circuits shown in (e). (g) Minimum decision time {\revision $\<T_{\rm dec}\>$ (Eq.~\ref{MFPT})} with a constraint on selectivity, solving $\min_{k_{\rm min}\leq k \leq k_{\rm max}} \{\langle T_{\rm dec} \rangle + \beta \log f\}$. Horizontal lines indicate the intervals between $\gamma/(\gamma+\omega(1+\gamma)/2)$ and $\gamma^2/(\gamma^2+\omega)$, respectively the minimum and maximum possible selectivity for the model shown in (b), for $\gamma=2$ and $8$ (purple and green) and $\omega=0.3$ and $3$ for $\gamma=8$, and $\omega=0.3$ and $1$ for $\gamma=2$ (increasing with color intensity). Black dots indicate the global minima for $\beta=0$. All simulations are done with $c_0=0.6$ nM and $k_{\rm max}=5$ s$^{-1}$. 
				\label{decoys}}}
	\end{center}
\end{figure*}

The cellular nucleus typically contains a mixture of transcription factors and other molecules which are not specific to the promoter's binding sites, but may still bind and initiate transcription (Fig.~\ref{decoys}a).
Assuming a promoter should be adapted to optimally encode information about its cognate ligand into its transcription activity, one may expect spurious binding to hurt that objective. 
This imposes constraints on how selective the transcription factor binding site may be, and how transcription factor concentrations may be distinguished in the face of additional noise coming from the binding of decoy binders.
Our next result shows that this is not always the case; under certain conditions promoters may make faster decisions while allowing for spurious binding, even at high concentrations of non-cognate ligands.

To model the presence of decoys, we go back to the single binding site, and add binding states to the Markov system corresponding to situations where a decoy is bound to the binding site instead of the cognate transcription factor (Fig.~\ref{decoys}b). The decoys bind and activate the gene with the same rates $k_{\rm b|on/off}$ as the cognate binder, but they unbind faster, with rates $k'_{\rm u,on/off}$. Decoys are present in concentration $c'$, which may be larger than the reference concentration of the transcription factor, $c_0$.

The relative abundance of the decoy is set by the ratio $\omega=c'/c_0$. The specificity of the binding site for its cognate transcription factor relative to the decoy is given by the ratio of unbinding rates $\gamma=k_{\rm u,on/off}'/k_{\rm u,on/off}$.
We numerically searched for networks with maximal $\dotF$ as a function of these two main parameters, $\omega$ and $\gamma$. Fig.~\ref{decoys}c shows the information rate $\dotF$ relative to the case of no decoy ($\omega=0$), at low $c_0$. Increasing the number of decoys (large $\omega$) hurts {\revision the precision of concentration estimation}; however, increasing specificity $\gamma$ mitigates this loss, because activation events triggered by decoy binding are then much shorter than by cognate binding, and
an ideal observer of the transcriptional activity can use information from the duration of activation events to sort out cognate activation events from decoy ones.

In all the optimal networks that we found numerically, the rate of activation in absence of binding ($k_{\rm on|u}$) was always best set to its minimal value $k_{\rm min}$, as these events carry no information about the ligands. Consequently, activation always occurs after binding of either the cognate or decoy ligand.
We define the selectivity $f$ as the fraction of time the system is active following a cognate binding event, relative to the total active time. As expected, this selectivity decreases with the amount of decoys, but increases with specificity, $\gamma$ (Fig.~\ref{decoys}d).

In a general equilibrium system, selectivity cannot be higher than $k_{\rm on}c/(k_{\rm on}c+k_{\rm on}'c')=\gamma/(\omega+\gamma)$.
Ninio and Hopfield \cite{Hopfield1974, Ninio1975} pointed out that selectivity could be improved in non-equilibrium circuits through kinetic proofreading. The idea is to force the system to clear multiple checkpoints, each of which gives an opportunity to test specificity (with efficiency $\gamma$), before reaching a productive state. In our circuits, the system allows for 2 such checkpoints: binding, and activation, each of which may be terminated specifically by unbinding, with rates $k_{\rm u,off}$ and $k_{\rm u,on}$. An architecture that optimizes these checkpoints can reach a maximal selectivity $f=\gamma^2/(\omega+\gamma^2)$ (see Appendix~\ref{App:selectivity}).
As a point of comparison, it is also useful to consider the simplest non-kinetic proofreading scheme in which all back rates are minimal. Note that while this network is highly non-equilibrium, it corresponds to the extension to decoys of the simplest irreversible loop of Fig.~\ref{difflimitted_simple}b. Its selectivity, $f=\gamma/(\omega(1+\gamma)/2+\gamma)$, is worse than the equilibrium one, because of the minimal amount of time $\sim k_{\rm max}^{-1}$ spent unbound but still active, even when activation was triggered by decoy binding.

We find that in general networks that are optimal for precision do not always make the most of the kinetic proofreading mechanism: plotting the realized selectivity $f$ in the optimal networks in Fig.~\ref{decoys}d, we find that it is far away from the maximally achievable one (dashed lines), especially if ligands are hard to discriminate ($\gamma=2$). However, optimal networks are still more selective than the simplest, non-proofreading network (dotted lines).

{\TT Different equilibria along the sensitivity-selectivity tradeoff are reflected in the architectures of optimal networks, which undergo two transitions with increasing decoy concentration (Fig.~\ref{decoys} e and f). These transitions are governed by two key rates: for large enough $\omega$, the back rate $k_{\rm u|off}$ activates the proofreading mechanism (blue shade in Fig.~\ref{decoys}f), which makes bursts caused by cognate and non-cognate binding more distinguishable. This comes at the expense of longer waiting times between activation bursts, and hence lesser throughput. Note that in this regime, unlike in Fig.~\ref{difflimitted_simple}, the Fisher rate does rely on ON times, because burst durations are informative about the identity of the ligand (cognate vs non-cognate). However, when $\omega \lesssim \gamma$, selectivity is high enough for the circuit to exploit an additional $c_0$-dependence of ON durations for concentration sensing, by switching on $k_{\rm b|on}$ (beige shade in Fig.~\ref{decoys}f, see Appendix~\ref{App:selectivity}). Although we find this rate saturates its diffusion limit ($4Da$) in optimal circuits and visibly affects $f$, its contribution to the Fisher rate is negligible when that rate is small compared to the deactivation rate $k_{\rm max}$.}

Conversely, networks with the largest selectivity are far from optimal in terms of precision, as measured by $\dotF$. This is illustrated in Fig.~\ref{decoys}g, which shows the best mean decision time, obtained from $\dotF$ through Eqs.~\ref{dotF} and \ref{MFPT} as a function of a fixed value of the selectivity $f$ (obtained by constrained optimization with Lagrange multipliers, see Appendix~\ref{App:lagrange}). The fastest decisions are obtained for relatively poor selectivity, relative to what is achievable through kinetic proofreading. The reason why optimal networks do not care much about selectivity is that decoy activations can be statistically distinguished from cognate ones, mitigating the harm of their interference. But kinetic proofreading does come at the cost of reducing the number of fruitful binding events leading to expression. Faced with this trade-off between purity of ligand and throughput, optimal circuits favor throughput.

\subsection{Quantifying irreversibility of optimal networks}
\label{results_noneq}

\begin{figure*}
	\begin{center}
		\includegraphics[width=\textwidth]{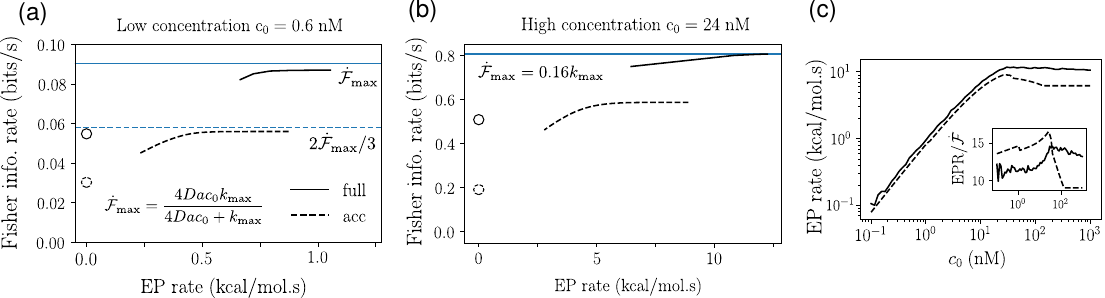}
		{\caption{Entropy production and circuit performance across decoding schemes. (a) Tradeoff between EPR and Fisher information rate for the full decoding scheme (solid) and accumulated activity decoding (dashed), at low concentration. {\revision Curves delimit Pareto fronts which we compute using the technique of Lagrange multipliers.} Circles indicate optimal equilibrium solutions. Increasing the allowable entropy production produces a discontinuous jump from equilibrium to non-equilibrium with no optima in between. Horizontal lines indicate upper limits of each decoding scheme. (b) Analogous plot at high concentration. (c) Entropy production rate at 90$\%$ of the maximum over a range of concentrations. Since the absolute maximum EPR depends closely on the lower bounds on rates, which are chosen for numerical convenience, we plot the value at the onset of the plateaus shown in (a) and (b). Inset: Entropy production rate per decision versus concentration. \label{optcircuit}}}
	\end{center}
\end{figure*}

All the optimal networks we have discussed so far are strongly out of equilibrium, as they typically follow irreversible paths across states, breaking microscopic time symmetry. How far a circuit is from equilibrium may be quantified rigorously through the entropy production rate:
\beq
\mathrm{EPR}=\sum_{i,j}  \pi_i q_{ij} \log{\frac{\pi_i q_{ij}}{\pi_jq_{ji}}},
\label{EPrate}
\eeq
where $q_{ij}$ denote the transition rate from state $i$ to $j$, and $\pi_i$ denotes the steady-state distribution of that system, which satisfies: $\sum_{j} (q_{ji}\pi_j -q_{ij}\pi_i)=0$. The EPR multiplied by $RT$ corresponds the minimal amount of energy per mol and per unit time (or power) that must be dissipated to maintain the observed irreversibility.

To assess the importance of irreversibility for precision, in Fig.~\ref{optcircuit}a and b we optimized simple networks for a single binding site with no decoys (Fig.~\ref{4circuit}a) for $\dotF$ and $\dotF_m$, but under a fixed EPR budget, using Lagrange multipliers, for a low (Fig.~\ref{optcircuit}a) and a high (Fig.~\ref{optcircuit}b) value of the reference concentration $c_0$. {\revision Note the Thermodynamic Uncertainty Relations (TUR) \cite{Puglisi2025}, which put bounds on the fluctuations of currents as a function of the entropy production rate, do not apply here since $\dotF$ is not a current.}

We see that irreversibility contributes substantially to the performance, as equilibrium networks (EPR=0, circles) perform 30\% to 50\% less well than the non-equilibrium ones. {\revision To evaluate the order of magnitude of the energy that is required to achieve near-maximum performance, we evaluate the EPR at which the Fisher information rate reaches $90\%$ of its maximal value. This EPR, which is represented as function of $c_0$ in Fig.~\ref{optcircuit}c, grows with concentration. It goes from vanishingly small consumptions at low $c_0$, to $\sim 10$ kcal/mol/s at large $c_0$.} {\revision The architecture of the optimal nonequilibrium circuit is given in Fig.~\ref{difflimitted_simple}b, while the optimal equilibrium circuit has non-vanishing rates for all reactions. Intuitively, optimal circuit architectures for decision allow the system to return to the OFF-unbound state as quickly as possible, while passing through the ON states, in which mRNA is produced. Going through this cycle produces entropy. Satisfying the equilibrium requirement means the return to the OFF-unbound state takes longer, allowing for fewer independent measurements of concentration and hence lower precision.}

At low concentrations, the gene gets activated rarely, with diffusion-limited rate $4Dac_0$, each time consuming the energy of one activation cycle, which explains the scaling of EPR with $c_0$. To obtain a normalized measure of dissipation that accounts for that, we plotted the ratio $\mathrm{EPR}/\dotF$, which is proportional to the amount of dissipation per decision (Fig.~\ref{optcircuit}c, inset). That normalized dissipation is in fact larger at small concentrations than at high ones, but it peaks at the transition concentration $c_0^*$ between the two regimes.

\section{Discussion}


Biological systems must often make decisions between discrete fates. An example from early fly development is the establishment of gap gene expression with single-cell resolution, in which small differences in TF concentration result in distinct readouts and gene expression programs~\cite{Gregor2007, Tran2018}. Given an unlimited amount of time, it is theoretically possible to distinguish any pair of concentrations as they will result in slightly different mean expression levels. However, development happens on relatively fast time scales, over which observations of gene expression activity are noisy because of the stochastic nature of transcription.
The SPRT approach developed in this paper had previously been applied to show that decisions about hunchback expression in the early development of the fly embryo can be made in under 3 minutes \cite{Desponds2020}. Those calculations assumed perfect coupling between binding and activation, ignoring the recruitment of the RNA polymerase. Modeling gene activation explicitly and separately from TF binding leads to a technical complication in the computation of the likelihood, due to the fact the durations of active and inactive periods are correlated through the TF binding state, which is hidden from the observer of gene activity. In this paper we showed how to compute the likelihoods necessary for implementing SPRT using dynamic programming. This in turn allowed us to predict the architecture of optimal networks for the task of distinguishing two nearby concentrations of the transcription factor.

We have shown that the information encoded in the full trace of activation and inactivation times is larger than the information encoded in the cumulated time of activation. This result formalizes and generalizes the previous observation that fundamental physical bounds on concentration sensing, first studied by Berg and Purcell \cite{Berg1977}, could be improved by a factor 2 by using maximum likelihood estimators \cite{Endres2009b}. To show this, we established a mathematical equivalence between physical bounds on concentration sensing and decision times under the SPRT. We showed that both are governed, in the limit of small concentration differences or equivalently large observation times, by the Fisher information rate $\dotF$. The difference between the two schemes is whether the observer has access to the whole history of gene activation, or just the cumulated activity.

This opens the question of how can cells access that historical information and decode the signal. Accumulating mRNA to approximate the total time of activation is one possible decoding mechanism~\cite{Lammers2023}. However, regardless of optimality arguments, we know that cells do not accumulate mRNA: mRNA has a short finite lifetimes, shorter than the division times of cells, even in prokaryotes. Decoding may be done at the next level---protein production, which may itself be subject to active post-transcriptional and post-translational regulations. When these proteins are themselves transcription factors, as it is the case for {\it hunchback} in fly development, decoding could happen further downstream, through {\revision the regulation of multiple downstream genes}. In sum, many mechanisms are available to process the raw information contained in the full temporal trace of gene activity, beyond the mere cumulated activation time. Some of these schemes were explored in Desponds et al~\cite{Desponds2020} for simple binding models. {\revision In fact, an ideal decoder that computes the log-likelihood ratio $\mathcal{R}(t)$ in the SPRT framework would only need to store as many variables as there are hidden nodes in the forward algorithm (see the Methods section), which in our model is just the number of binding sites + 1. This is true even for full-trajectory decoding; there is never any need to store historical data about mRNA activity.} It would be interesting to explore how these mechanisms could be used to make efficient decisions downstream of the optimal networks that we predict.

Our work shares concepts and tools with that of Lammers et al~\cite{Lammers2023}, who optimized networks for distinguishing nearby concentrations using SPRT. However, there are 3 key differences with our approach, which affects the predicted form of the optimal networks. First, we optimized for the SPRT over the full trace, which contains more information than the cumulated activity, as already discussed, but also changes the form of the predicted optimal networks and corresponding responses. 
Second, in \cite{Lammers2023} information rates are normalized by the cycle time, i.e. the typical time it takes to complete an activation event. By contrast, we optimized raw rates expressed in bits per second. Then, without further constraints, the optimal network is always to send all rates to arbitrarily large values to increase information rate. {\revision Since rates are finite and often diffusion limited, these schemes  lead to unphysical solutions.} This brings us to the third difference: we imposed upper bounds on rates of binding, unbinding, activation and inactivation. Physically, binding rates are limited by the availability of transcription factors, and by the speed with which they find their binding site on the DNA through diffusion. Unbinding rates cannot be arbitrarily large if the TF is to bind specifically to DNA. Bounds on (in)activation rates correspond to the minimum bound time required to trigger downstream mechanisms, such as activating a gene, recruiting RNAP, or shutting it off \cite{Desponds2020}. As it turns out, these {\revision constraints} are key to shaping the response properties of optimal networks. For instance, we saw that, at low concentrations, optimal networks aim to complete transcriptional bursts as quickly as possible within those constraints, making them inactive most of the time, in contrast to the results of \cite{Lammers2023}, in which it is optimal to be at half of the maximum activity.  {\revision The other effect of constraints on rates other than binding is that they limit the ability of non-equilibrium 4-state circuits to approach equivalent equlibrium 2-state circuits, for which precision bounds were previously derived \cite{Berg1977,Endres2009}}.
Another difference is that, unlike Ref.~\cite{Lammers2023}, we find that optimal networks do not have sharp response functions.

{\revision In the presence of decoys, we find that optimal circuits generically use kinetic-proofreading mechanisms to discriminate cognate from non-cognate ligands, consistent with previous theoretical analyses~\cite{CepedaHumerez2015}. However, discriminability is only one of the contributor of precision, and optimal circuits balance it against throughput. It may be beneficial to initiate transcription even upon binding the wrong ligands, if it allows for processing more binding events. This is an example of the classical speed-accuracy trade-off, where in our case the role of accuracy is played by specificity. A recent study, in the very different context of DNA replication \cite{Ravasio2025}, argued that this trade-off may be inverted in the high error regime, thanks to the stalling effect: kinetic proofreading could then evolve from selecting on replication speed only. In the regime of very low mutation rates however, the usual accuracy-speed trade-off is still predicted to apply. Since there is no stalling in our setting, we find no equivalent effect. Note that, unlike for DNA replication where speed is defined irrespectively of accuracy, in our context speed implies information rate. It is intrinsically linked to the presence of the correct ligand, meaning that the analogy with DNA replication is very limited.}

We find that optimal networks are generically out of equilibrium, and that imposing constraints on dissipation (quantified by entropy production) degrades precision. {\revision Microscopic analyses of the transcription dynamics at a single gene copy of yeast have shown that the regulatory process for transcription is cyclic and irreversible~\cite{Shelansky2024}. Functioning out of equilibrium was suggested as a mechanism for reconciling fast regulatory kinetics and high regulatory specificity.} Past work has tried to address how limiting entropy production constrains optimal circuits~\cite{Mancini2016, Szymanska-Rozek2019, Desponds2020,Lammers2023}. However, it worth recalling that the energetic costs incurred by that dissipation are dwarfed by those of mRNA transcription, which involves 2 phosphate bond hydrolyses per added nucleotide, i.e., of the order of thousands of kcal/mol per transcribed mRNA, not counting the costs of synthetizing nucleotides. By contrast, the dissipation caused by irreversible effects in our optimal networks is of the order of 10 kcal/mol at best. This suggests that while being out of equilibrium helps, the amount of energy required to realize that improvement is negligible. This discrepancy between the information-theoretic costs of computation and actual metabolic costs also highlights the fact that the amount of dissipation depends on the level of coarse graining~\cite{Szymanska-Rozek2019, Crisanti2012, Puglisi2010, Busiello2019}. Our simple models do not account for the full complexity of the processes of gene transcription, and may thus underestimate the true dissipation. Conversely, it is worth noting that, when coarse-grained, our models may yield very different amounts of energy dissipation. For instance, the irreversible loop of binding/activation/unbinding/deactivation that we find in the simplest setup may be reduced, when (in)activation is fast, to a simpler system with 2 state\,---\,bound/active, unbound/inactive\,---\,which has no dissipation at all.

Our optimization for precision makes several predictions that can be compared to data. 
We find that, when the binding rate of TF to the DNA is limiting, $4Dac_0\ll k_{\rm max}$, optimal networks should remain minimally active by completing each activation cycle as soon as possible, in order to be available for new informative binding events. This prediction holds even when considering an observer that only has access to the cumulated activity time. {\revision However, when kinetic constraints on the rates of other processes are large, so that diffusion is no longer the main limit, the model predicts that genes should be relatively well expressed at the decision point.}
In the early stages of fly development, genes are substantially active at positions along the anterior-posterior axis where cell fate is decided \cite{Lucas2013,Zoller2018}. Our theory would then suggest that the TF binding rate is not the only limiting factor, but that the time necessary to complete each activation cycle, $\sim k_{\rm max}^{-1}$, also limits precision. It also predicts that, in presence of multiple binding sites, it is only beneficial to use cooperativity when the binding rates are not limiting ($4Dac_0\sim k_{\rm max}$). The fact that the response to maternal gradients is often cooperative \cite{Gregor2007,Dubuis2013,Tran2018} further indicates that TF binding rates may not be the main limiting factor to precision, but rather that they are balanced by other rates.

{\revision Other biochemical constraints may add to the effect of limits on biochemical rates.}
A limitation of our approach, also shared by Ref.~\cite{Lammers2023}, is the assumption that mRNA production tracks gene activity deterministically. In reality, mRNA transcription is discrete, so that the number of mRNA products is a linear but noisy function of the activity time. While this approximation is acceptable in the regime of reasonably high gene transcription activity, it breaks down when activation periods are short, which is often the case in our optimal networks. With transcription rates of $15-30$ mRNAs per minute \cite{Zoller2018}, an optimally fast activation cycle of duration $\sim 2k_{\rm max}^{-1}< 0.5s$ would produce at most one mRNA transcript per binding event. Transcription would then be very stochastic, which would limit the benefit of short activation cycles predicted by our theory. A possible way to account for this effect would be to constrain the rate of deactivation to very low values, to ensure that each activation event leads to at least a few transcripts, so that $k_{\rm max}\sim 0.25 - 1$s$^{-1}$ (instead of 4s$^{-1}$ as in the figures). An interesting future direction would be to explore the impact of transcription stochasticity on precision and on the corresponding optimal networks.

Another prediction of our theory is the form of the regulation function, and more precisely how on and off times depend on gene activity (Fig.~\ref{difflimitted_simple}c). Experimental data on the gap genes in the fly
\cite{Zoller2018,Chen2024} show both on and off times depend on the gene activity, but in a way that their harmonic mean is constant. This is consistent with our predictions, provided that we consider the cumulated activity, rather than the full trace, as the output used by the decoder. {\revision We take the inverse of ON and OFF times as a proxy for the ON and OFF rates reported experimentally in \cite{Zoller2018,Chen2024}}.

More broadly,  the question of how cells respond to environmental signals and exploit information can be broken down into two parts: how they encode and store information about the input and how they decode it. Decoding matters, yet the way information is stored already can limit the ability of cells to exploit information and respond. Additionally, different types of cellular hardware may be better suited for specific encoding schemes, as they evolve together. Our results show that physical and structural constraints play an important role in possible encoding schemes in signaling circuits.

\section{Methods}

\subsection{Likelihood of expression traces}\label{sec:likelihood}
To compute decision time, we need to estimate numerically the Fisher information rate defined in \eqref{dotF}. In this section we give an outline of how to perform that computation.

The system explores a set of states $\sigma\in\mathcal{A}\bigcup\mathcal{I}$ where $\mathcal{A}=\{A_i,i=1,\ldots,n\}$ are the active states (transcription is on), and $\mathcal{I}=\{I_i,i=1,\ldots,n\}$ the corresponding set of inactive states (transcription is off). The transition rates between state $\sigma$ and $\sigma'$ are encoded in the matrix $Q(c)=(q_{\sigma',\sigma}(c))$. The likelihood for a particular trajectory of states, where the system spends $t_1$ in state $\sigma_1$, $t_2$ in state $\sigma_2$, etc, reads:
\beq\label{proba}
\begin{split}
  P((t_1,\sigma_1),(t_2,\sigma_2),\ldots |c) &=e^{-t_1\sum_{\sigma}q_{\sigma,\sigma_1}(c)} q_{\sigma_2,\sigma_1}(c)\\
  &\times e^{-t_2\sum_{\sigma}q_{\sigma,\sigma_2}(c)}q_{\sigma_3,\sigma_2}(c)\cdots
  \end{split}
\eeq
However, this sequence of visited states is not accessible to the transcription state of the system. Instead, the output of the system is the temporal trace of transcriptional activity {\revision (Fig.~\ref{4circuit}c)}, which can be encoded by a sequence of times spent in the active and inactive states, $T^-_1,T^+_1,T^-_2,\ldots$, where $T^-_1=\sum_{i\in s^-_{1}}t_i$, $T^+_1=\sum_{i\in s^+_1}t_i$, and so on, where $s^-_1$ is the sequence of inactive states first visited before visiting any active state, $s^+_1$ the following sequence of active states visited before being inactivated again, $s_2^-$ the sequence of inactive states visited after that, etc.

We need to compute the likelihood of this sequence of residence times for a given concentration $c$:
\beq\label{likelihood}
  P(O|c)=P(T^-_1,T^+_1,T^-_2,\ldots|c).
\eeq
The likelihood is a sum of the probabilities \eqref{proba} of all possible scenarios $\{t_i,\sigma_i\}$ that are consistent with the output $(T^-_1,T^+_1,T^-_2,\ldots)$.
The difficulty is that these durations are correlated with each other, through the particular realizations of the states through which the transitions between active and inactive states happen. Therefore, the likelihood in \eqref{likelihood} cannot be written as a product over those transitions.

We can make progress by keeping some, if not all, information about the visited hidden states, namely, only the states to which the system transitions when it switches between active and inactive states. Call $\sigma_i^-$ the first visited state of $s_i^-$,
$\sigma_i^+$ the first state of $s_i^+$,
etc. Then the likelihood of the trace augmented by that information does factorize:
\beq~\label{factorized_likelihood}
\begin{split}
  &P(\{T_i^-,T_i^+,\sigma_i^-, \sigma_i^+\})\\
  &\quad =P(\sigma_1^-)\prod_{i\geq 1} P(T_i^-,\sigma_i^+|\sigma_i^-) P(T_i^+,\sigma_{i+1}^-|\sigma_i^-),
\end{split}
\eeq
where $P(T_i^-,\sigma_i^+|\sigma_i^-)$ is the probability that, starting at inactive state $\sigma_i^-$, it takes time $T_i^-$ for the system to reach its first active state, and that that active state is $\sigma_i^+$; and similarly for $P(T_i^+,\sigma_{i+1}^-|\sigma_i^-$). Note that, with that additional conditioning on the entry state, waiting times are now uncorrelated, so that probabilities may be factorized.
Then the likelihood \eqref{likelihood} may be rewritten as:
\beq
P(O|c)=\sum_{\sigma_1^-,\sigma_1^+\sigma_2^-,\ldots}P(\{T_i^-,T_i^+,\sigma_i^-, \sigma_i^+\}).
\eeq

This sum involves $2^{2n}$ terms, where $2n$ is the number of transitions between active and inactive states, however it can be computed efficiently using dynamic programming approaches.
To do this, we define intermediate, conditional sums up to the $k^{\rm
  th}$ activation cycle:
\begin{eqnarray}
  Z^+_{k, \sigma_{k}^+}(c)&=&P(T_1^-,T_1^+,\ldots,T_{k}^-, \sigma_{k}^+|c),\\
  Z^-_{k, \sigma_{k}^-}(c)&=&P(T_1^-,T_1^+,\ldots,T_{k-1}^+, \sigma_{k}^-|c).                       
\end{eqnarray}
Then we have the recursion:
\begin{eqnarray}
  Z^+_{k, \sigma_k^+}(c)&=&\sum_{\sigma_{k}^-} P(T_k^-,\sigma_k^+|\sigma_k^-) Z^-_{k, \sigma_{k}^-}(c)\label{rec1}\\
    Z^-_{k, \sigma_{k}^-}(c)&=&\sum_{\sigma_{k-1}^+} P(T_{k-1}^+,\sigma_k^-|\sigma_{k-1}^+) Z^+_{k-1, \sigma_{k-1}^+}(c),\label{rec2} \quad
\end{eqnarray}
and the final likelihood is obtained as the sum over the final
($n^{\rm th}$) state:
\beq
P(O|c)=\sum_{\sigma_n^+} Z^+_{n,\sigma_n^+}(c).
\eeq

The conditional probabilities for the on and off durations in Eqs.~\eqref{rec1}-\eqref{rec2} may be computed
by first-passage time distribution techniques \cite{Kemeny1960a}.
To recall, $P(T_k^-,\sigma_k^+|\sigma_k^-)$ is the probability density
of the $k^{\text{th}}$ period of inactivity $T_k^-$, the time taken
for the state to reach an active state $\sigma_k^+$ for the first
time, starting from inactive state $\sigma_k^-$. It can be obtained by
building a matrix $Q^-(c)=(q^-_{\sigma',\sigma}(c))$ in which all
rates in the matrix $Q(c)$ that transition out of the active states
$(A_i)$ are set to zero, making $\cal A$ an absorbing set,
\begin{equation}
	q^-_{\sigma', \sigma}(c) = 
	\begin{cases}
		0 & \text{if } \sigma' \in \mathcal{A}\text{ and } \sigma \in \mathcal{I}, \\
		q_{\sigma', \sigma}(c) & \text{otherwise.}
	\end{cases}
\end{equation}
The transition probability matrix of the system thus obtained reads
\begin{equation}
  P^-(t, c) = \exp[tQ^-(c)]. \label{trans-matrix-absorbing}.
\end{equation}
The density of the first passage time $T_k^-$ can then be read off
from entries of the time derivative of \eqref{trans-matrix-absorbing}, 
\begin{equation}\label{eq:firstpass}
	P(T_k^-,\sigma_k^+|\sigma_k^-) = \big(Q^-(c) \exp[tQ^-(c)]\big)_{\sigma_k^-,\sigma_k^+}.
\end{equation}
The same procedure can be used, setting the inactive states as
absorbing, to obtain $P(T_{k-1}^+,\sigma_k^-|\sigma_{k-1}^+)$.

\subsection{Numerical optimization}

Although computing the likelihood of a single trace can be done efficiently as described above, our study aims to optimize the SPRT drift, Eq. \eqref{V_lim}. This is a trajectory average over observable traces, or, since it is a self-averaging quantity, an estimate with a single observation trace of long enough duration. Obtaining  small estimate variances for the purposes of optimization remains computationally expensive for generic circuits, but necessary, since one cannot rule out that correlations between bursts might contain information about concentration.

We carried out the optimizations of figures 2-5 in two stages. First, the full trajectory average in Eq. \eqref{V_lim} was estimated by Monte Carlo sampling with a Julia script, using a single simulated trace $\tilde{O}(t)$ for a large $t$ 
\beq
\dot{\cal F}\simeq \frac{1}{t}\log \frac{P(\tilde{O}(t)\vert c)}{P(\tilde{O}(t)\vert c+\delta c)}.
\eeq

This was in turn maximized numerically using the Differential Evolution algorithm from Julia's Optimization package. 

We then observed that $\dot{\cal F}$-maximizing circuits always verify $k_{\rm on|u}=k_{\rm min} \simeq 0$, which aligns with the intuitive fact that a transcript initiated without TF binding does not carry any information about its concentration. In a second iteration, we repeated all optimizations imposing this equality, whose consequence is that an ON/OFF burst always starts in the (unbound, OFF) state.
In that case, the likelihood may be factorized as a product over each activation cycle:
\beq\label{likelihood}
  P(O|c)=\prod_{i}P(T^-_i,T^+_i|c).
\eeq
Each term of this product may be computed exactly by using \eqref{eq:firstpass} with $\sigma_i^-=$unbound, and summing over $\sigma^+_i=$bound cognate and bound non-cognate. This procedure doesn't require any recursive matrix product or Monte Carlo, and is differentiable.

We ran a second round of optimizations with the Fisher information rate instead computed as an exact double integral over ON and OFF durations, yielding a computational speedup of over 2 orders of magnitude:
\beq
\dot{\cal F}=\frac{c_0^2}{\tau}\int_0^\infty \int_0^\infty \frac{\left (\partial_{c_0} P(T^-, T^+\vert c_0)\right )^2}{P(T^-, T^+\vert c_0)}dT^-dT^+
\eeq

This speedup was key in enabling us to explore the effects of varying $\omega$ and $\gamma$, in addition to concentration.

\subsection{Joint Fisher information across cognate and non-cognate concentrations}

The decoy-prone circuit architectures of Figure \ref{decoys} encode information about both $c_0$ and $c'$, the concentration of decoy ligands, in the mRNA transcript. In these circuits, measuring the sensitivity to $c_0$ using the univariate formula for Fisher information \eqref{dotF} amounts to assuming that $c'$ is known perfectly to a downstream observer. Exploiting this knowledge could lead to a better estimation of $c_0$, which is unrealistic in real receptors. To rule out such solutions, we modeled the joint uncertainty in both parameters using the inverse Fisher information matrix,
\beq
\mathbf{C} = \begin{bmatrix}
	\dot{\cal F}_{c_0c_0} && \dot{\cal F}_{c_0c'}\\
	\dot{\cal F}_{c_0c'} && \dot{\cal F}_{c'c'}
\end{bmatrix}^{-1},
\eeq\\
with the cross terms given by
\beq
\dot{\cal F}_{c_0c'} = \frac{c_0c'}{\tau}\int_0^\infty \int_0^\infty \frac{\partial_{c_0}P(c_0,c')\partial_{c'}P(c_0,c')}{P(c_0,c')}dT^+dT^-.
\eeq
The results of Figure \ref{decoys} are then obtained by maximizing
\beq
\dotF_{c_0}=\frac{1}{[\mathbf{C}]_{c_0c_0}},
\eeq
which corresponds to the Fisher information rate for $c_0$ after accounting for the uncertainty on $c'$.

~\label{App:Methods}

\section*{Acknowledgements}

The study was supported by Agence Nationale de la Recherche grant no
ANR-22-CE95-0005-01 ``DISTANT,'' and by the CZI Theory Initiative.

\bibliographystyle{pnas}

\onecolumngrid

\newpage

\appendix

\renewcommand{\thefigure}{S\arabic{figure}}
\newcommand{\la}{\langle}
\newcommand{\ra}{\rangle}
\newcommand{\mc}{\mathcal}
\newcommand{\tx}{\text}
\setcounter{figure}{0}

\section{Bounds on Fisher information rate for optimal 4-state circuits (without decoys)}
\label{App:V_limits_4s}

Our numerical results establish that in the absence of decoys, $\dot{{\cal F}}$-maximizing circuits come in one of two architectures, depending on the decoding scheme. They are shown in the insets of Figure \ref{difflimitted_simple} b and d. Here we examine the Fisher rate of these architectures in detail, shedding light on the limits of $\dot{{\cal F}}$ from Figure \ref{difflimitted_simple} (b) at low and high concentration, for accumulated and full activity decoding. 

At a given Bicoid concentration $c_0$, the parameters of the circuit are the binding rates, $k_{\rm b|off}c_0$ and $k_{\rm b|on}c_0$, the rate of unbinding during active transcription $k_{\rm u|on}$, the rate of activating once Bicoid is bound, $k_{\rm{on|b}}$, and finally the rate of deactivating once Bicoid unbinds, $k_{\rm{off|u}}$. The constraint imposed by diffusion sets $k_{\rm{b|\{on,off\}}} \leq 4 a D$, with $a$ the linear size of the receptor and $D$ the diffusion constant of Bicoid. All other rates are assumed to take values smaller than $k_{\rm max}$ and we impose a lower limit of $10^{-5}$ on all rates for numerical stability in simulations.

\subsection{Low concentration regime}

The diffusion-limited regime is characterized by few binding events, their rates being small in comparison to the rates of activating and deactivating transcription. This limit simplifies the analysis of multi-state circuits, even when computing likelihood ratios of full transcription activity. At low concentration, the Fisher rate of an optimal 4 state receptor approaches that of a two-state receptor that activates as soon as a ligand binds, and deactivates when it unbinds. In that limit $k_{\rm{on|b}} \gg k_{\rm{b|off}}c_0$, and OFF times are distributed as $P(T_{\rm{off}}) \sim e^{-k_{\rm b|off}c_0 T_{\tx{off}}}$. The distribution of ON times, is a convolution of two exponentials, but this complexity is bypassed by noting that ON times are independent of concentration in the optimal architecture, and therefore do not contribute to the estimator.

With accumulated decoding however, a 4-state circuit has a 4/3 improvement in sensitivity relative to a two-state architecture, and does not reduce to it even at low concentrations. The general expression of the reduced Fisher rate  \cite{Lammers2023} is given by $\dotF_m=s^2/\dot v$, where $\dot v=\lim_{t\to\infty}\text{Var}(T_{\rm on}(t))/t$.

Fluctuations in $T_{\rm on}(t)$ come from binding and activation events. This was studied in \cite{Mora2010} for a two-state receptor, but their results hold for our setting in the low concentration limit, when times of inactivity are approximately exponentially distributed with rate $k_{\rm{b|off}}c_0$ (see Eq. S63 in \cite{Mora2010}, supplemental information):
\begin{equation}
	\label{mrna_vars}
	\dot v = [1 + \langle (\delta T^+)^2 \rangle / \langle T^+\rangle^2 ]\frac{k_{\rm{b|off}}c_0/\langle T^+\rangle}{\big(1/\langle T^+\rangle + k_{\rm{b|off}}c_0\big)^3}.
\end{equation}

The ON binding rate $k_{\rm{b|on}}$ is negligible compared to $k_{\rm{off|u}}$ at low concentration. The mean time spent ON then reads
\begin{equation}
	\langle T^+\rangle \simeq 1/k_{\rm{u|on}} + 1/k_{\rm{off|u}},
\end{equation}
and the coefficient of variation of ON time is a sum of the CVs of two exponential waiting times
\begin{equation}
	\langle (\delta T^+)^2 \rangle / \langle T^+\rangle^2 = \frac{(k_{\rm{u|on}})^2 + (k_{\rm{off|u}})^2}{(k_{\rm{u|on}} + k_{\rm{off|u}})^2}.
\end{equation}
It reaches a minimum of 1/2 when $k_{\rm{u|on}} = k_{\rm{off|u}}=k_{\rm max}$, making the leftmost factor on the RHS of \eqref{mrna_vars} equal to 3/2.

Besides we have
\begin{equation}
	s^2 = \pi_{\text{on}}^2(1-\pi_{\text{on}})^2={\left[\frac{k_{\rm b|off}c_0/\langle T^+\rangle}{(1/\langle T^+\rangle + k_{\rm b|off}c_0)^2}\right]}^2,
\end{equation}

Putting it all back into $\dotF_m=s^2/\dot v$, we get:
\begin{equation}
	\dot{{\cal F}}_{m} = \frac{2}{3} \frac{k_{\rm b|off}c_0/\langle T^+\rangle}{1/\langle T^+\rangle + k_{\rm b|off}c_0} \simeq 2k_{\rm b|off}c_0/3 = 2\dot{{\cal F}}/3.
\end{equation}

\subsection{High concentration regime}

Our numerical results indicate that above a precise threshold of concentration $c_0$, the optimal OFF binding rate $c_0k_{b|\rm off}$ is constant with respect to $c_0$ for MLE decoding, and both ON and OFF binding rates are constant for BP decoding, as is the Fisher rate. This places the circuit far from any limits that enable 2-state simplifications, and so a full analysis of the 4-state circuit is required. In this section we perform this analysis for the full trajectory decoding scheme.

We begin by noting that for optimal circuits shown in Figure~\ref{difflimitted_simple}, the bound inactivation rate $k_{\rm off|b}$ and the unbound activation rate $k_{\rm on|u}$ saturate their lower bounds, rendering their effect negligible. This makes activation and inactivation times $T_{i}^+$ and $T_i^-$ independent in optimal circuits, and greatly simplifies the likelihood:
\begin{align}
	P\left(\{T_i^-,T_i^+\}_{i=1}^n\,|\,c\right) = \prod_{i=1}^n P_-&(T_i^-)P_+(T_i^+),\label{pjoint} \quad \quad \\
	P_-(t) = \frac{ax}{x-a}\Big (e^{-at} - e^{-xt} \Big),& \label{poff} \quad \quad \begin{cases}
		x = c_0k_{b|\rm off}\\
		a = k_{\rm on|b} \\
	\end{cases} \\
	P_+(t) = \frac{bc}{s_2 - s_1}\Big (e^{-s_1t} - e^{-s_2t}& \Big),\quad \, \begin{cases}
		b = k_{u|\rm on}  \\
		c = k_{\rm off|u} \\
	\end{cases}\label{pon} \\
	s_{\sfrac{1}{2}} = \frac{1}{2}\left(-y-b-c \;\footnotesize{\sfrac{+}{-}}\; \sqrt{\Delta}\right),& \quad \quad \begin{cases}
		\Delta = (y+b+c)^2-4bc  \\
		y = c_0k_{b|\rm on}
	\end{cases} 
\end{align}

As a result of independence, the Fisher information per burst cycle splits into two contributions,
\begin{equation}
	{\cal F}_{\rm cyc}(c) = -\left \langle \frac{\partial^2\log P_-(t)}{(\partial \ln c_0)^2}\right \rangle_c  -\left \langle \frac{\partial^2\log P_+(t)}{(\partial \ln c_0)^2}\right \rangle_c= {{\cal F}}_{\rm off}(c) + {{\cal F}}_{\rm on}(c). \label{von-voff-eq}
\end{equation}
Moreover, ${{\cal F}}_{\rm on}(c)$ is negligible in optimal architectures, since ON times are independent of concentration. To obtain the first term, the second derivative of the logarithm of~\eqref{poff} reads
\begin{align}
	\frac{\partial^2 \log P_-(t)}{(\partial \ln c_0)^2} &= -1 +  \left(\frac{x}{x-a}\right)^2 - e^{(x+a)t}\left(\frac{xt}{e^{xt} - e^{at}} \right)^2 ,
\end{align}
from which the Fisher information of an OFF time is obtained by averaging over $P_-(t)$:
\begin{align}
	{\cal F}_{\rm off}(c_0) &= 1 - \left(\frac{x}{x-a}\right)^2 + \frac{ax}{x-a}\int_0^{\infty} e^{(x+a)t}\Big (e^{-at} - e^{-xt} \Big)\left(\frac{xt}{e^{xt} - e^{at}} \right)^2 dt, \\
	& = 1 - \left(\frac{x}{x-a}\right)^2 + \frac{ax^3}{x-a}\int_0^{\infty} \frac{t^2}{e^{xt} - e^{at}}  dt, \\
	& = 1 - \left(\frac{x}{x-a}\right)^2 + \frac{ax^3}{x-a}\left [\frac{2}{(x-a)^3}\zeta\left(3,\max\left \{\frac{x}{x-a}, \frac{a}{a-x}\right\}\right)\right],\\
	& = 1 - \alpha^2 +\frac{2a}{x}\alpha^4\zeta\left(3,\beta\right), \label{fisher_off}
\end{align}
where $\alpha = \frac{x}{x-a}$ and $\beta = \alpha$ if $x>a$, and $\beta = 1-\alpha$ if $x<a$. 

We numerically evaluated~\eqref{fisher_off} 
in the simulations discussed in the main text. From there, maximizing $\dot{{\cal F}}(c_0) = {\cal F}_{\rm cyc}(c_0)/\tau$ over $x,y,a,b$, and $c$ yields
\begin{align}
	\dot{{\cal F}}^*(c_0)&= \min \left\{4aDc_0,\gamma_1 \,k_{\rm max}\right\}, \\
	x^* &= \min \left \{ 4aDc_0,\gamma_2 \,k_{\rm max}\right\},\\
	y^* &= k_{\rm min},\\
	a &= b = c = k_{\rm max},\\
	\gamma_1 &\simeq 0.161649, \\
	\gamma_2 &\simeq 0.503427.
\end{align}

\section{Proofreading architectures of the decoy-prone receptor}
\label{App:selectivity}

\begin{figure}[ht]
	\includegraphics[width=\textwidth]{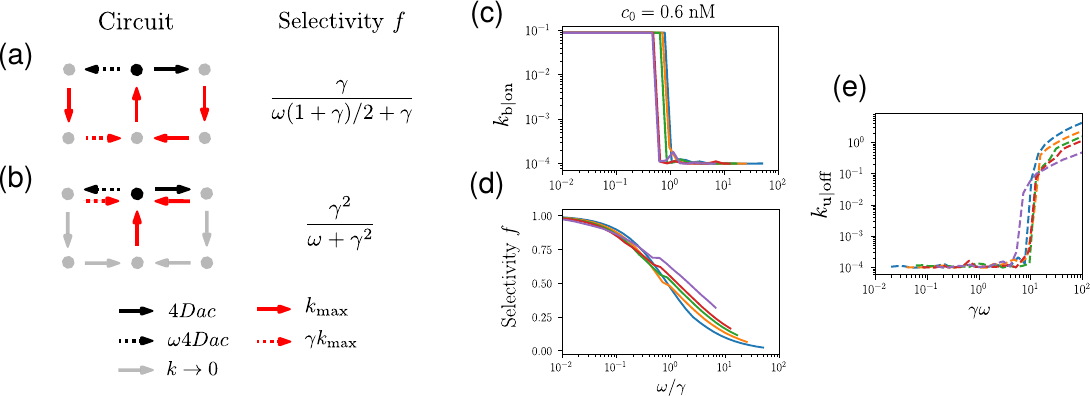}
	\caption{Minimum and maximum selectivity (a,b) Circuit architectures with minimum (a) and maximum selectivity (b). The ON binding rates (c) and selectivity (d) collapse when plotted against $\omega/\gamma$, hinting that the ON-mediated proofreading mechanism turns off when selectivity is below half, coinciding with $\omega \simeq \gamma$. Colors represent $\gamma \in \{2,4,6,8,15\}.$ \label{proofreading}}
\end{figure}

The ability of a receptor to initiate transcription due to the correct ligand (selectivity) in the presence of decoys is limited by its architecture. Two types of proofreading schemes emerge from optimizing the Fisher rate at various decoy concentrations and specificities. 

The minimum and maximum selectivity circuits are shown in Fig.~\ref{proofreading} (a) and (b), respectively. We define selectivity as the ratio of time spent transcribing due to a correct binding event. This time includes the residence time in state (on, cognate bound), and the time spent in (on, unbound) weighed by the probability of entering that state due to a cognate binding event.

The upper and lower bounds on $f$ are straightforward to obtain. For (a),
\beq
f=\frac{2/k_{\rm max}}{2/k_{\rm max}+\omega(1/k_{\rm max}+1/\gamma k_{\rm max})}=\frac{\gamma}{\omega(\gamma+1)/2+\gamma},
\eeq
In (b), the ideal proofreading scheme makes the event of entering a transcribing state due to decoy binding weigh $\omega/\gamma$ relative to cognate binding, instead of $\omega$. This limit of maximum selectivity is only achieved when $k_{\rm on|b}, k_{\rm u|on} \to 0$, where the Fisher information is also zero. In addition, time spent in the (on, decoy bound) state has weight $1/\gamma$ relative to (on, cognate bound). This leads to odds ratio of finding the system in the active state with the cognate ligand rather than the decoy to be $\gamma^2/\omega$, leading to $f=\gamma^2/(\omega+\gamma^2)$.

When selectivity is high, ON burst times depend strongly on TF concentration; an optimal circuit is therefore incentivized to use them as an information source for sensing $c_0$. This is mediated by $k_{\rm b|on}$, which activates this concentration at a transition point around $f \simeq 0.5$, which corresponds to $\omega \simeq \gamma$ (Figure~\ref{proofreading} c and d).

Moreover, we find that OFF-mediated proofreading becomes active when $\gamma\omega$ crosses a threshold (Figure~\ref{proofreading} e). OFF times carry more information about concentration when burst time statistics differ sufficiently between cognate and decoy binding events. This is controlled primarily by $\gamma$. At high decoy concentrations, optimal circuits also have an incentive to proofread to reduce the total time spent transcribing due to the wrong binding events. 

\section{Optimization with constraints on selectivity and entropy production rate}
\label{App:lagrange}

To explore the sensitivity-selectivity tradeoff directly we solve the following constrained optimization problem:
\beq
\max_{k_{\rm min}\leq k_{\rm b} \leq 4aD, k_{\rm min}\leq k_{\rm \{u,on,off\}} \leq k_{\rm max}} \dot{{\cal F}}(c_0) + \beta \log f
\eeq
Solutions are represented in Figure~\ref{decoys} g, with $\dot{{\cal F}}$ substituted by the SPRT decision time by applying Eq.\eqref{MFPT}, at fixed concentration $c_0$ and sweeping the tradeoff parameter $\beta$ from 0 to $\tan(\pi/2 - 0.05) \simeq 20$.

The tradeoff between sensitivity and entropy production rate in Figure~\ref{optcircuit} was obtained with
\beq
\max_{k_{\rm min}\leq k_{\rm b} \leq 4aD, k_{\rm min}\leq k_{\rm \{u,on,off\}} \leq k_{\rm max}} \dot{{\cal F}}(c_0) - \beta \text{EPR}
\eeq
for $\beta$ ranging from $10^{-7}$ to 0.06 in logarithmic scale.

\end{document}